\documentstyle[multicol,aps,prl,epsf]{revtex}

\begin{document}

\title{Experimental Measurement of the Persistence Exponent of the Planar
Ising Model}

\author{B. Yurke, A.N. Pargellis, S.N. Majumdar$^{1*}$, C. Sire$^2$}

\address{
Bell Laboratories, Lucent Technologies, Murray Hill, NJ 07974\\
$^1$ Physics Department, Yale University, New Haven, CT 06520-8120 \\
$^2$ Laboratoire de Physique Quantique, Universit\'e Paul Sabatier, 
31062 Toulouse Cedex, France}
\maketitle
\begin{abstract}
Using a twisted nematic liquid crystal system exhibiting planar Ising model
dynamics, we have measured the scaling exponent $\theta$ which characterizes
the time evolution, $p(t) \sim t^{-\theta}$, of the probability $p(t)$ that
the local order parameter has not switched its state by the time $t$. For 0.4
seconds to 200 seconds following the phase quench, the  system exhibits
scaling behavior and, measured over this interval, $\theta = 0.19 \pm 0.031$,
in good agreement with theoretical analysis and numerical simulations.

\vskip 0.25cm
\noindent PACS:  82.20.Fd, 02.50.Ey, 05.40.+j, 05.50.+q
\end{abstract}


\begin{multicols}{2}

There has been a recent surge of interest in determining the so-called
nontrivial ``persistence'' exponent $\theta$
\cite{breath,DBG,BDG,Stauffer,MH,DHP,DER,DOS,Boston,MS,MSBC} associated with
the dynamics of a phase ordering system following a quench from the
high-temperature phase to zero temperature. This exponent describes the
asymptotic power-law decay, $p(t) \sim t^{-\theta}$, of the probability that
the local order parameter $\phi({\bf x,t})$ has not changed sign up to time
$t$ after the quench. For the Ising model, $p(t)$ is simply the fraction of
spins that have not flipped up to time $t$, or, equivalently, the probability
that no interface has ever crossed a given spin. While there have been
considerable theoretical and numerical efforts directed towards calculating
$\theta$, so far there has been no experimental measurement of this exponent
in spin-like systems. A kind of persistence exponent was first introduced and
measured in a breath figure experiment \cite{breath}, and recently for soap
bubbles \cite{soap} (in this latter example $\theta$ takes a rather trivial
value, which can be inferred on simple physical grounds \cite{soap}). In this
Letter, we present the first experimental measurement of the nontrivial
persistence exponent for an Ising-like system, that we find in good agreement
with a recent theoretical calculation.

This power-law decay of persistence is quite ubiquitous and not just
restricted to the phase ordering dynamics of the Ising or Potts models. For
instance, a similar question arises in the study of Gaussian processes.
Considering such a process $m(t)$, the calculation of the probability that
this Gaussian walker never crosses the origin (or changes sign) is a
particularly difficult problem when this process is not Markovian
\cite{MS,MSBC}. For example, even for the simple diffusion equation,
$\partial_t\phi = \nabla^2\phi$, the probability that the Gaussian variable
$\phi(x,t)$ does not change sign up to time $t$, decays algebraically with $t$
with a nontrivial exponent $\theta$ \cite{MSBC}. This exponent $\theta$ also
appears in other contexts such as reaction-diffusion systems
\cite{KBR,Cardy,Howard} and driven diffusive systems \cite{CB}.
For a quench to
the critical point of spin systems, the persistence exponent $\theta$
associated with the total magnetization (another non-Markovian Gaussian
variable) has recently been argued to be a new non-equilibrium critical
exponent \cite{MBCS}.

For the purpose of the experimental results presented below, we will restrict
ourselves to the phase ordering dynamics of the Ising model at $T=0$. The
Ising model at $T=0$ has two ordered states: either all spins are up or they
are all down. Following a quench from the high-temperature initial state where
all spins are randomly up or down, the system tries to order locally. However,
since the symmetry between the two ordered phases is not broken by the quench
process, the two ground states compete with each other. As a result, domains
of both phases form, grow with time, and the system exhibits coarsening. At
late stages of coarsening, the morphology of the growing domains seems to
exhibit a self-similar structure in time once all length scales are rescaled
by $L(t)$, where $L(t)$ represents the typical linear dimension of a growing
domain. The scaling hypothesis also predicts that the equal-time two-point
correlation function has a simple scaling form, $\langle \phi({\bf 0}, t)
\phi({\bf r}, t)\rangle \sim g[r/L(t)]$, where the calculation of $g(x)$ has
been the subject of intense theoretical efforts  \cite{Review}. Also, the
length scale $L(t)$ grows algebraically with time, $L(t)\sim t^{1/2}$.
This growth law can be simply understood by noting that the
motion of the domain walls between the opposite phases is purely curvature
driven \cite{Review}.

The equal-time correlation function, however, only gives information about the
spatial structure of the system at a given time. To obtain information about
the temporal evolution, one useful measure is the unequal-time autocorrelation
function, $C(t,t')= \langle \phi({\bf r}, t') \phi({\bf r}, t)\rangle $. For
$t\gg t'$, this autocorrelation decays as $C(t,t')\sim
[L(t)/L(t')]^{-\lambda}$, where $\lambda$ was argued to be a new universal
nontrivial non-equilibrium exponent \cite{FH}. For the $d=2$ Ising model,
Fisher and Huse give a heuristic argument leading to $\lambda=5/4$ \cite{FH},
in good agreement with their numerical simulations. The experimental
measurement of $\lambda$ was carried out for the first time by
Mason et al.\cite{mason93}, using a twisted nematic liquid crystal sample
bounded by two glass plates (as will be explained in detail below). The
experimental value of $\lambda$ in $d=2$ was in very good agreement with the
theoretical prediction \cite{mason93}.

The important question, however, is whether this autocorrelation is sufficient
to characterize the full temporal evolution of the system. The answer is
presumably no, since the stochastic evolution process of a given spin is not
Gaussian and not Markovian. What is the minimal set of quantities needed to
specify the full distribution of this complex spatial and temporal
structure? It is extremely difficult to answer this general question as
the evolution of the order parameter field in general satisfies a nonlinear
partial differential equation. In the absence of such information, one
therefore looks for simple, easily measurable quantities that still give
important information about the history of the evolution process. One such
simple and natural quantity is the ``persistence'', i.e., the probability
$p(t)\sim t^{-\theta} \sim [L(t)]^{-2\theta}$ that a given spin does not flip
up to time $t$.

This fraction of unflipped spins $p(t)$, though relatively simple to measure
in numerical simulations \cite{BDG,DER,Stauffer,MS}, proves extremely hard to
compute theoretically.
The temporal evolution
of an individual spin is a non-Gaussian, non-Markovian process, and naturally,
any history dependent nonlocal quantity is very hard to compute for such
processes. Indeed, this persistence probability is not simply a two-point
correlation function of the initial and final times, but involves all
intermediate times as the process is, in general, non-Markovian.
Nevertheless, for the $d=1$ Ising model with Glauber dynamics at $T=0$, this
nontrivial exponent $\theta$ was recently computed exactly by Derrida et al.
\cite{DHP}, who found $\theta=3/8$. Unfortunately, their method cannot be
extended to higher experimentally relevant dimensions. For the $d=2$ spin flip
dynamics of the Ising model, only numerical estimates of the exponent $\theta
\approx 0.22$ \cite{BDG,Stauffer,DOS,MS} have been available to date.

An approximate analytical
method has recently been developed\cite{MS} to compute $\theta$
in any dimension for the deterministic nonlinear
Landau-Ginzburg equation, which describes the $T=0$ dynamics of an $isotropic$
spin system in the $continuum$. It involves the following few steps. First,
one assumes that the late-time dynamics of the discrete Ising spins are
correctly described by the continuum noiseless model-A equation for the
nonconserved order parameter field $\phi({\bf r},t)$. In fact, for the
experimental system discussed below, this continuum equation of motion for the
field is certainly more appropriate \cite{cornell}. Next, one makes a
nonlinear transformation $\phi({\bf r},t)=\sigma(m,t)$, where $m({\bf r},t)$
is an auxiliary field locally representing the distance to the nearest domain
wall \cite{MAZ,LM} and $\sigma$ is the equilibrium profile of a domain wall.
$\sigma$ has a sigmoid shape, which, in practice can be treated as the sign
function. One then assumes that $m({\bf r},t)$ is a true Gaussian field (an
approximation becoming exact in the large $d$ limit \cite{MAZ,LM}) which
allows the calculation of the two-point correlators of both $m$ and $\phi$
self-consistently. This method was originally introduced by Mazenko
\cite{MAZ,LM}, and proves very successful in calculating the exponent
$\lambda$ or the equal-time correlator $g(x)$. For example, in $d=2$, this
approximation yields $\lambda\approx 1.289$ to be compared with $1.25\pm 0.01$
obtained in simulations \cite{FH}. Once the above approximation has been made,
the probability of not flipping a spin is the same as the probability that the
Gaussian process $m$, at a given point in space, does not change sign up to
time $t$, as the equilibrium interface profile is an odd function
($\phi=\sigma (m)\sim\text{sign}(m)$). Still, the computation of the
probability of no zero crossing for the Gaussian process remains extremely
hard due to its non-Markovian nature.

However, the problem of computing $\theta$ can be mapped onto a quantum
mechanics problem \cite{MS} of sorts. $\theta$ is then found to be the
difference between two ground-state energies: that of a quantum-like problem
with a hard wall at the origin (this constraint results from the condition
that the Gaussian walker has to remain on, say, the positive axis), and the
other for the same system but without the wall. In these terms, the quantum
problem associated with a Markovian process is simply an harmonic oscillator
with frequency $\lambda/2$, where $\lambda$ is again the autocorrelation
exponent. One can then develop standard variational and perturbative methods
to compute the exponent $\theta$ for a process close enough to a Markovian
process. In $d=1$, our approximate theory yields $\theta \approx 0.35 $
\cite{MS} as compared with the exact value $\theta=3/8$ \cite{DHP}. For $d=2$,
we found $\theta \approx 0.19$ \cite{MS}. This last result is also consistent
with the estimate $\theta \approx 0.186$ obtained for $d=2$ in the
asymptotically exact $d\to\infty$ limit of Mazenko theory using the
independent intervals approximation for the zeros of the process $m(t)$
\cite{MSBC}, and in recent direct simulations of the time-dependent
Ginzburg-Landau (TDGL) equation (model-A) \cite{cornell}. Note that if the
process $m(t)$ were Markovian, one could show \cite{MS} that the simple
equality $\theta=\lambda/2\approx 0.63$ holds. The discrepancy with the actual
value is an indication of strong memory effects in this system.

For comparison with theory, liquid crystals afford the possibility of
designing experiments for which $\theta$ can be easily measured. Indeed, a
number of liquid crystal systems, possessing a variety of dimensionalities and
types of order parameters, have been used to measure various scaling exponents
characterizing the coarsening process. Such measurements have included the
increase in the characteristic scaling length, $L(t)$, in
three-dimensional\cite{chuang91,yurke92,snyder92,wong92} and
two-dimensional\cite{orihara86,nagaya87,shiwaku90,pargellis92,pargellis94}
systems. The twisted nematic system, exhibiting Ising-like behavior and
employed here for the measurement of $\theta$, has been previously used by
Orihara and co-workers \cite{orihara86,nagaya87}  to study the behavior of the
scaling  length $L(t)$ and, in addition, by Mason et al.\cite{mason93} to
measure $\lambda$. These studies used a liquid crystal sample placed between
two glass plates whose surfaces had been treated so that, after a thermal
quench from the isotropic phase to the nematic phase, the liquid crystal
organized itself into domains in which the director was forced to twist
clockwise or counterclockwise by $\pi/2$ in going from the surface of one
plate to the other. The boundary between two domains of opposite twist
consists of a twist disclination line. Regions of opposite twist correspond to
Ising model domains in which the spins all point up or all point down. The
system relaxes viscously, driven by the line tension of the disclination
lines.

The experimental apparatus used for the $\lambda$ measurements has been
described in detail by us previously\cite{mason93} and the data collected for
this report was extracted from the video tapes obtained in these earlier Ising
system studies. Briefly, the sample cell consists of a layer of liquid
crystal, $20 \mu$m thick, sandwiched between two parallel glass microscope
slides. The liquid crystal used was
trans-(trans)-4-methoxy-4'-$n$-pentyl-1,1'-bicyclohexyl (Merck, CCH-501, or
equivalently  ZLI-3005), with an isotropic-to-nematic phase transition of 37
$^{o}C$. The molecular orientation at each slide's surface was forced to be in
the plane of the surface and oriented unidirectionally. The surfaces of the
glass plates were prepared by first dipping them into a $0.1 \%$ by weight
solution of polyvinyl alcohol and then buffing the surface in one direction
with a soft cloth. The two slides were mounted orthogonally to each other,
thus defining a square region of liquid crystal, about 2.5 cm on a side. The
orthogonal orientation of the plates with respect to one another forces the
nematic liquid crystal to phase separate into two domains, as the molecular
orientation undergoes either a clockwise or counterclockwise twist of $\pi/2$
radians as one goes from one plate to the other. The circular dichroic effect
was used to make the domains of opposite twist visible. A square region, 1.5
mm on a side, was observed using a Nikon E Plan x4 objective and recording the
images with a high-speed color recording system (NAC, HSV-400). Images were
recorded every 5 ms, each frame labeled at the top with the run number and
time since the recording began.

For a well percolated system undergoing self-similar coarsening, the ratio of
the area occupied by domains of one particular twist to the total area as a
function of time should remain constant and have the value 0.5. This ratio,
averaged over the 15 runs discussed here, was 0.375 at the time of the quench,
reaching within only 0.2 sec the value of $0.576 \pm 0.016$, at which it
remained for the 200 second time interval over which  data was recorded. There
are several possible reasons for the deviation of this ratio from the value of
0.5. The region between domains over which the director  field is distorted
from the pure clockwise or counterclockwise twist is comparable to the
thickness of the cell, 20 microns. This makes the optical image of the
boundary between domains somewhat ill-defined. Hence, optical effects could
cause us to preferentially choose one domain over the other when deciding
whether a given point belongs to a region of clockwise or counterclockwise
twist. This would be particularly serious at early times and probably accounts
for the low value (0.375) of the ratio measured at early times. The ratio at
late times is 15\% higher than the expected value of 0.5 and may be due to a
small bias in the domain nucleation, domain growth favoring domains of one
particular twist, and, at very late times, small sample statistics. For
example, if alignment of the upper and lower plate deviate from perfect
orthogonality one twist direction will be favored over the other
\cite{nagaya87}.

\begin{figure}[h]
\narrowtext
\epsfxsize=\hsize
\epsfbox{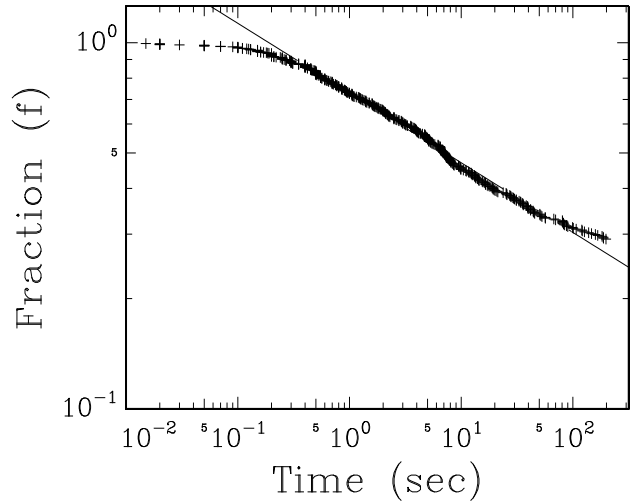}
\caption{Probability $p(t)$ versus time that an individual point in the system
has not switched its state at least once by time $t$. The solid line is a
least-squares fit to the data and has a slope of 0.19.}
\end{figure}

We measured, for each of 40 points (in an ordered, rectangular array) in each
of 15 runs \cite{footnote1}, the time $t$ after the quench when the molecular
orientation first switched from one twist direction to the other. The contrast
was sufficient to enable us to distinguish the orientation at a particular
point immediately after the phase transition. Fig. 1 is a log-log plot showing
the scaling of the probability $p(t)$ an individual point has not switched its
orientation by time $t$. A least-squares fit to the data in the range 0.4 to
200 seconds gives a slope, $\theta = 0.190 \pm 0.031$, shown by the solid
line. The initial approach to scaling is most likely due to the initial
nonzero size of the scale length $L$ at the time of the quench and the
identification bias resulting from the diffuse nature of the optical image of
the domain boundaries. This initial scale length is on the order of the
thickness of the cell ($20 \mu$m), resulting in a minimum domain size at the
time of the quench. Also noticeable at late times is a ``tail'', of reduced
slope, in $p(t)$. The tail at late times could be due to several different
effects: pinning of domain walls in the sample immediately outside of the
region observed or nonorthogonal orientation of the glass plates (giving rise
to a preferred molecular twist). The wiggles in the data are finite sample
size effects reflecting the history of how macroscopic regions changed their
orientation in particular runs.

In conclusion, our measured value of the persistence exponent  $\theta = 0.190
\pm 0.031$ is in good agreement with the theoretical estimate $\theta \approx
0.19$ for an $isotropic$ bidimensional spin system. 

We thank A.J. Bray, S.J. Cornell, K. Damle, S. Sachdev, and T. Senthil for
fruitful discussions. The research of one of us (S.N.M.) was supported by
NSF Grant No. DMR-92-24290. Laboratoire de Physique Quantique is Unit\'e Mixte
de Recherche C5626 of Centre National de la Recherche Scientifique (CNRS).


\end{multicols}


\begin{references}
\bibitem[*]{} permanent address after Oct. 96: Tata Institute of Fundamental
Research, Theoretical Physics Group, Homi Bhabha Road Mumbai, Bombay 400005,
India.

\bibitem{breath} M. Marcos-Martin, D. Beysens, J.-P. Bouchaud, C. Godr\`eche
and I. Yekutieli, {\sl Physica A} {\bf 214}, 396 (1995).

\bibitem{DBG} B. Derrida, A.J. Bray, and C. Godr\`eche, 
{\sl J. Phys. A} {\bf 27}, L357 (1994).

\bibitem{BDG} A.J. Bray, B. Derrida and C. Godr\`eche, {\sl Europhys.
\ Lett.} {\bf 27}, 175 (1994).

\bibitem{Stauffer} D. Stauffer, {\sl J. Phys.\ A } {\bf 27}, 5029 (1994).

\bibitem{MH} S.N. Majumdar and D.A. Huse, {\sl Phys. Rev. E} {\bf 52},
270 (1995).

\bibitem{DHP} B. Derrida, V. Hakim and V. Pasquier, {\sl Phys. Rev. Lett.} 
{\bf 75}, 751 (1995). 

\bibitem{DER} B. Derrida, {\sl J. Phys. A} {\bf 28}, 1481 (1995).

\bibitem{DOS} B. Derrida, P.M.C. de Oliveira and D. Stauffer, 
{\sl Physica A} {\bf 224}, 604 (1996). 

\bibitem{Boston} E. Ben-Naim, L. Frachebourg and P.L. Krapivsky, 
{\sl Phys.  Rev. E} {\bf 53}, 3078 (1996).

\bibitem{MS} S.N. Majumdar and C. Sire, {\sl Phys. Rev. Lett.} {\bf 77}, 1420
(1996); C. Sire, S.N. Majumdar and A. R\"udinger, to be published.

\bibitem{MSBC} S.N. Majumdar, C. Sire, A.J. Bray and S.J. Cornell, {\sl Phys.
Rev. Lett.} {\bf 77}, 2867 (1996); B. Derrida, V. Hakim and R. Zeitak, {\sl
Phys. Rev. Lett.} {\bf 77}, 2871 (1996).

\bibitem{soap} W.Y. Tam, R. Zeitak, K.Y. Szeto and J. Stavans, to be
published; B. Levitan and E. Domany, to be published; for soap bubbles
$(d>1)$, one simply has $\theta=\lambda/2=d/2$; see C. Sire and S.N. Majumdar,
{\sl Phys. Rev. E} {\bf 52}, 244 (1995).

\bibitem{KBR} P.L. Krapivsky, E. Ben-Naim and S. Redner, {\sl Phys. Rev. E}
{\bf 50}, 2474 (1995).

\bibitem{Cardy} J. Cardy, {\sl J. Phys. A}, {\bf 28}, L19 (1995).

\bibitem{Howard} M. Howard, {\sl J. Phys. A}, {\bf 29}, 3437 (1996).

\bibitem{CB} S.J. Cornell and A.J. Bray, {\sl Phys. Rev. E}
{\bf 54}, 1153 (1996).

\bibitem{MBCS} S.N. Majumdar, A.J. Bray, S.J. Cornell and C. Sire, {\sl Phys.
Rev. Lett.} {\bf 77}, 3704 (1996); see also D. Stauffer, {\sl Int. J. Mod.
Phys. C} in press.

\bibitem{Review} For a recent review on the kinetics of phase ordering,
see A.J. Bray, {\sl Advances in Physics} {\bf 43}, 357 (1994).

\bibitem{FH} D.S. Fisher and D.A. Huse, {\sl Phys. Rev. B} {\bf 38}, 373
(1988).

\bibitem{mason93}
N. Mason, A.N. Pargellis and B. Yurke,  {\sl Phys. Rev. Lett.} {\bf 70}, 190
(1993). 

\bibitem{cornell} S.J. Cornell, unpublished. Note that the values of $\theta$
obtained in $d=2$ Ising simulations are systematically slightly higher than
those obtained for the noiseless TDGL. This puzzling fact may be an indication
that these two systems are not in the same universality class as far as
measuring $\theta$ is concerned (see \cite{MS} for a discussion in $d=3$).

\bibitem{MAZ} G.F. Mazenko, {\sl Phys. Rev. Lett.} {\bf 63}, 1605 (1989).

\bibitem{LM} F. Liu and G.F. Mazenko, {\sl Phys. Rev. B} {\bf 44}, 9185
(1991).

\bibitem{chuang91}
I. Chuang, R. Durrer, N. Turok and B. Yurke, {\sl Science} {\bf 251}, 1336
(1991); I. Chuang, N. Turok and B. Yurke, {\sl Phys. Rev. Lett.} {\bf 66}, 2472
(1991). 

\bibitem{yurke92} B. Yurke, A.N. Pargellis, I. Chuang and N. Turok, {\sl
Physica B} {\bf 178}, 56 (1992). 

\bibitem{snyder92} R. Snyder, A.N. Pargellis, P. Graham and B. Yurke, {\sl
Phys. Rev. A} {\bf 45}, R2169 (1992). 

\bibitem{wong92} A.P.Y. Wong, P. Wiltzius and B. Yurke, {\sl Phys. Rev.
Lett.} {\bf 68}, 3583 (1992).

\bibitem{orihara86}
H. Orihara and Y. Ishibashi, {\sl J. Phys. Soc. Jpn.} {\bf 55}, 2151 (1986).

\bibitem{nagaya87}
T. Nagaya, H. Orihara and Y. Ishibashi, {\sl J. Phys. Soc. Jpn.} {\bf 56},
3086 (1987).

\bibitem{shiwaku90} T. Shiwaku, A. Nakai, H. Hasegawa and T. Hashimoto,  {\sl
Polymer Commun.} {\bf 28}, 174 (1987); {\sl Macromolecules} {\bf 23}, 1590
(1990).

\bibitem{pargellis92}
A.N. Pargellis, P. Finn, J.W. Goodby, P. Panizza, B. Yurke and P.E. Cladis,
{\sl Phys. Rev. A} {\bf 46}, 7765 (1992). 

\bibitem{pargellis94}
A.N. Pargellis, S. Green and B. Yurke, {\sl Phys. Rev. E} {\bf 49}, 4250
(1994). 

\bibitem{footnote1}
Forty points were used in all except one run where an imperfection in one
corner pinned the domain wall. In this case, the 12 points were eliminated
that were in the immediate vicinity of the pinning site.



\end{references}
\end{document}